\documentclass[preprintnumbers,showpacs,amsmath,amssymb,floatfix,prd,nofootinbib,showkeys]{revtex4-1}
\usepackage{graphicx}
\usepackage{bbm}
\usepackage{bm}
\usepackage{amsfonts}
\usepackage{epstopdf}
\usepackage[colorlinks,linkcolor=red,anchorcolor=blue,citecolor=green]{hyperref}
\usepackage{subfigure}
\usepackage{float}
\usepackage{enumerate}
\usepackage{fix-cm}
\newcommand\diff{\mathrm{d}}

\newcommand\e{\mathrm{e}}

\begin{document}
\title{
	Solar system tests for realistic $f(T)$ models with nonminimal torsion-matter coupling
}
\author{Rui-Hui \surname{Lin}}
\author{Xiang-Hua \surname{Zhai}}
\email[]{zhaixh@shnu.edu.cn}
\author{Xin-Zhou \surname{Li}}
\email[]{kychz@shnu.edu.cn}
\affiliation{Shanghai United Center for Astrophysics (SUCA), Shanghai Normal University,
100 Guilin Road, Shanghai 200234, China}
\begin{abstract}

In the previous paper,
we have constructed two $f(T)$ models with nonminimal torsion-matter coupling extension,
which are successful in describing the evolution history of the Universe
including the radiation-dominated era,
the matter-dominated era,
and the present accelerating expansion.
Meantime, the significant advantage of these models is that
they could avoid the cosmological constant problem of $\Lambda$CDM.
However, the nonminimal coupling between matter and torsion will
affect the tests of Solar system.
In this paper,
we study the effects of Solar system in these models,
including the gravitation redshift, geodetic effect and perihelion preccesion.
We find that Model I can pass all three of the Solar system tests.
For Model II,
the parameter is constrained
by the measure of the perihelion precession of Mercury.

\end{abstract}
\maketitle
\section{Introduction}
General Relativity (GR)
uses the metric tensor as the fundamental dynamical variable,
and chooses the torsionless Levi-Civit{\'a} connection to describe gravitation field.
In Teleparallel Equivalence of General Relativity (TEGR)
\cite{Hayashi1979,Aldrovandi2012,Maluf2013},
the equivalent gravitation theory constructed by Einstein,
the curvatureless Weizenb{\"o}ck connection is chosen
instead,% the Levi-Civit{\'a} one's stead,
and tetrad field plays the role of fundamental variable.
TEGR provides the gauge structure of gravitation:
the gauge theory for the group of translation on the tangent bundle of the spacetime.
In this sense,
teleparalellism may be considered as an approach
to include gravity into the unification of gauge theories,
and maybe further, to quantize the gravity.

Under the framework of TEGR,
different modification of gravity theory has been proposed
for better understanding of the currently accelerating universe\cite{Amanullah2010,Suzuki2011}.
One possible scheme is $f(T)$ theory,
in which the torsion scalar $T$ in the gravitation Lagrangian
is replaced by an arbitrary function $f(T)$
\cite{Ferraro2007,Bengochea2009,Dong2013,Feng2014,Junior2016}.
Compared to $f(R)$ theory,
the field equations are second order instead of fourth one,
which is an important advantage of $f(T)$ theory.
As a further extension,
and an analogy to that of $f(R)$ theories\cite{Bertolami2007,Sotiriou2008,Harko2010,Thakur2011,Xi2012,Harko2015a,Harko2015,Xu2016},
nonminimal torsion-matter coupling $f(T)$ gravity
has been proposed and studied\cite{Harko2014}.
In the previous research\cite{Feng2015},
using the observation data of type Ia supernovae(SNeIa),
cosmic microwave background(CMB), and baryon acoustic oscillations(BAO),
we established two concrete $f(T)$ models with nonminimal torsion-matter coupling extension,
and found that they are successful in describing the
observation of the Universe and its large-scale structure and evolution.
The joint fitting led to
$\Omega_\text{m0}=0.255\pm0.010$,
$\Omega_\text{b0}h^2=0.0221\pm0.0003$
and $H_0=68.54\pm1.27$ for model I and
$\Omega_\text{m0}=0.306\pm0.010$,
$\Omega_\text{b0}h^2=0.0225\pm0.0003$
and $H_0=60.97\pm0.44$ for model II
at $1\sigma$ confidence level,
where $H_0$ is the Hubble parameter at present,
$h=H_0/(100\text{km/s/Mpc})$,
and $\Omega_\text{m0}$ and $\Omega_\text{b0}$ are
the density parameters of dark matter and baryon matter.

On the other hand,
any modified gravity theory must confront the Solar system tests which have been passed in GR\cite{Will1993,Kagramanova2006,Will2014}.
In fact, Solar system effects in minimal coupling $f(T)$ models
have been considered\cite{Iorio2012,Xie2013,Farrugia2016}.
In Ref.\cite{Farrugia2016},
the authors have investigated the perihelion precession,
light bending, Shapiro time delay and gravitation redshift
in minimal coupling $f(T)$ gravity with $f(T)=T+\alpha T^n$,
and constrained the parameter $\alpha$ for $n=2,\:3$.
However, nonminimal geometry-matter coupling theories predict that
the movements of massive particles are no longer geodesic,
which will have influences on some of the Solar system effects.
The similar cases have been studied in some nonminimal coupling $f(R)$ theories
(see e.g. \cite{Bertolami2007,Harko2011,Bertolami2013}).

In this paper,
we consider the Solar system tests including
the gravitation redshift, geodetic effect, and perihelion precession
in the two models we have constructed in Ref.\cite{Feng2015}.
The gravitation redshift is one representative effect that
does not involve massive particle.
Since the two models we are considering assume that
the coupling between radiation and torsion is still minimal for simplicity,
this kind of effects is practically the same as the minimal coupling theory.
The geodetic effect and perihelion precession, however,
involve the movement of massive objects,
and thus should be affected by the nonminimal coupling between matter and torsion.
Judging from this,
it is necessary to combine the cosmological restriction with Solar system effects
for the parameters in two models.
We have gotten the cosmological restriction using the observation data of SNeIa, CMB and BAO as follows\cite{Feng2015}:
$A=0.188\pm 0,048$, $B=0.510\pm0.060$ (for Model I);
$A=0.633\pm0.012$ and $B$ is a free parameter (for Model II).
It is our task in this paper to check
whether or not these values of parameters
will pass the Solar system tests.

The paper is organized as follows:
In Section \ref{review}
we briefly review the $f(T)$ theory
with power law torsion-matter coupling extension
and our two cosmologically fitted models.
In Sect. \ref{bgmetric},
we calculate the weak field limit of
the spherically symmetric background metric.
In Sects. \ref{gr}, \ref{ge}, and \ref{pp}
we consider the gravitation redshift,
geodetic effect and perihelion precession
in our two models, respectively.
We conclude our research in Sect. \ref{sum}.

We are going to use the Latin alphabet
$(a,b,c,\cdots=0,1,2,3)$
to denote the tangent space indices,
and Greek alphabet
$(\mu,\nu,\rho,\cdots=0,1,2,3)$
to denote the spacetime indices.
We assume the Lorentz metric of Minkowski spacetime
\begin{equation}
	\eta=\eta_{ab}\diff x^a\otimes\diff x^b
	\label{eta}
\end{equation}
has the form
$\eta_{ab}=\text{diag}(+1,-1,-1,-1)$.
And we use the unit
$c=8\pi G=1$
throughout the paper.

\section{Brief review of the theory}
\label{review}
On the spacetime differentiable manifold $\mathcal{M}$,
one can generally find dual pairs of linearly independent frame fields $\{e_a,e^a\}$,
such that $e^a(e_b)=\delta^a_b$.
One particular pair would be
the gradients $\{\partial_a\}$ of the tangent space coordinates $\{x^a\}$
and their covectors $\{\diff x^a\}$.
One such frame field, also known as the tetrad or vierbein field,
forms a base for the vectors on the tangent space
$T_p\mathcal{M}$ at each point $p\in\mathcal{M}$.
On the overlap area of different patches of $\mathcal{M}$,
the members of one base can be written in terms of members of the other,
e.g.
\begin{equation}
	e_a=e_a^{\:\mu}\partial_\mu\quad\text{or}\quad e^a=e^a_{\:\mu}\diff x^\mu.
	\label{bases}
\end{equation}

The spacetime metric $g$
\begin{equation}
	g=g_{\mu\nu}\diff x^\mu\otimes\diff x^\nu
	\label{metricg}
\end{equation}
is related to the tangent space metric $\eta$ by
\begin{equation}
	\eta_{ab}=g(e_a,e_b)=g_{\mu\nu}e_a^{\:\mu}e_b^{\:\nu},
	\label{metrictrans}
\end{equation}
or conversely,
\begin{equation}
	g_{\mu\nu}=\eta_{ab}e^a_{\:\mu}e^b_{\:\nu}.
	\label{metrictrans1}
\end{equation}
And hence the determinant
\begin{equation}
	|e|\equiv\text{det}(e^a_{\:\mu})=\sqrt{-g}.
	\label{metricdet}
\end{equation}

TEGR uses these vierbein fields as dynamical variables to define
the Weitzenb{\"o}ck connection\cite{Weitzenbock1923}
\begin{equation}
	\tilde{\Gamma}_{\nu\mu}^{\lambda}\equiv e_a^{\:\lambda}\partial_\mu e_{\:\nu}^a=-e_{\:\nu}^a\partial_\mu e_a^{\:\lambda}.
	\label{connection}
\end{equation}
The torsion and contorsion tensors are then given by
\begin{equation}
	T_{\:\:\mu\nu}^\lambda\equiv\tilde{\Gamma}_{\nu\mu}^\lambda-\tilde{\Gamma}_{\mu\nu}^\lambda=e_i^\lambda(\partial_\mu e_\nu^i-\partial_\nu e_\mu^i),
	\label{torsiontensor}
\end{equation}
\begin{equation}
	K^{\mu\nu}_{\quad\rho}\equiv-\frac12\left(T^{\mu\nu}_{\quad\rho}-T^{\nu\mu}_{\quad\rho}-T_{\rho}^{\:\mu\nu}\right).
	\label{contorsion}
\end{equation}
Utilizing the tensor
\begin{equation}
	S_\rho^{\:\mu\nu}\equiv\frac12\left(K^{\mu\nu}_{\quad\rho}+\delta_\rho^\mu T^{\lambda\nu}_{\quad\lambda}-\delta_\rho^\nu T^{\lambda\mu}_{\quad\lambda}\right),
	\label{auxtensor}
\end{equation}
one can define the torsion scalar
\begin{equation}
	T\equiv T^\rho_{\:\mu\nu}S_\rho^{\:\mu\nu}.
	\label{torsionscalar}
\end{equation}
TEGR\cite{Hayashi1979,Aldrovandi2012,Maluf2013} uses this scalar $T$ as the gravitation Lagrangian density.

For another vierbein field $\{e'_a,e'^a\}$,
contracting Eq.\eqref{metrictrans1} with $e'^{\:\mu}_ce'^{\:\nu}_d$,
one has
\begin{equation}
	g_{\mu\nu}e'^{\:\mu}_ce'^{\:\nu}_d=\eta_{cd}=\eta_{ab}\left(e^a_{\:\mu}e'^{\:\mu}_c\right)\left(e^b_{\:\nu}e'^{\:\nu}_d\right).
	\label{lorentz}
\end{equation}
This means $\Lambda^a_{\:c}=e^a_{\:\mu}e'^{\:\mu}_c$ is one of the entries of a Lorentz transformation.
That is, different vierbein fields satisfying the Lorentz transformation relation
\begin{equation}
	e'^c_{\:\:\mu}=\Lambda^c_{\:a}e^a_{\:\mu}
	\label{lorentz1}
\end{equation}
will give the same spacetime metric.
Though this does not make much difference in TEGR,
it does have some significant effects
on $f(T)$ theories,
one of which is that the $f(T)$ theories are no longer Lorentz invariant theories\cite{Li2011a}.
The fact that different vierbein fields (related by a Lorentz group) lead to the same metric
calls for an appropriate choice of vierbein for $f(T)$ theories and their extensions,
especially when spherical coordinates is considered\cite{Tamanini2012}.

One of the extention of $f(T)$ theories is the
non-minimal torsion-matter coupling $f(T)$ gravity.
Proposed by Harko et al.\cite{Harko2014},
the action of the non-minimal torsion-matter coupling $f(T)$ gravity can be written as
\begin{equation}
	S=-\frac12\int|e|\left[1+f_1(T)\right]T\diff^4x+\int|e|\left[1+f_2(T)\right]\mathcal{L}_\text{m}\diff^4x,
	\label{action}
\end{equation}
where $\mathcal{L}_m$ is the Lagrangian density of matter.
For simplicity, the evolution of radiation is assumed to be
the same as that in TEGR or GR,
and the coupling between radiation and geometric is assumed to be minimal.
To include radiation in the action
(to consider e.g. early time evolution of the universe),
one can straightforwardly append the radiation term to Eq.\eqref{action}.
With the action principle applied on Eq.\eqref{action}
with respect to the vierbein,
the equation of motion is then given by
\begin{equation}
		\frac4{|e|}f\partial_\beta(|e|S_\sigma^{\:\:\alpha\beta}e_a^{\:\sigma})+4e_a^{\:\sigma}S_\sigma^{\:\:\alpha\beta}\partial_\beta f+4fS_\rho^{\:\:\alpha\sigma}T^\rho_{\:\:\sigma\beta}e_a^{\:\beta}+(1+f_1)Te_a^{\:\alpha}=-2(1+f_2)\mathcal{T}_\beta^{\:\alpha}e_a^{\:\beta},
	\label{eom}
\end{equation}
where $f=f(T)\equiv1+f_1(T)+f_1'(T)T-2f_2'(T)\mathcal{L}_\text{m}$, and $\mathcal{T}_\beta^{\:\alpha}$
is the energy-momentum tensor of matter given by
\begin{equation}
	\frac{\delta(|e|\mathcal{L}_\text{m})}{\delta e^a_{\:\alpha}}=-|e|\mathcal{T}_\beta^{\:\alpha}e_a^{\:\beta}.
	\label{emtensor}
\end{equation}
And it takes the usual form for perfect fluid
\begin{equation}
	\mathcal{T}_{\mu\nu}=pg_{\mu\nu}-(\varepsilon+p)u_\mu u_\nu,
	\label{emtensor1}
\end{equation}
where $p$ and $\varepsilon$ are the pressure and energy density of the matter, respectively,
and $u^\mu$ is the 4-velocity.

The covariant derivative (related to the Levi-Civit{\'a} connection) of Eq.\eqref{eom} gives
\begin{equation}
	\nabla^\nu\mathcal{T}_{\mu\nu}=-\frac{f_2'(T)\nabla^\nu T}{1+f_2(T)}(\mathcal{T}_{\mu\nu}+g_{\mu\nu}\mathcal{L}_\text{m}).
	\label{covder}
\end{equation}
This suggests that the energy-momentum tensor is no longer conservative.
However,
contracting Eq.\eqref{covder} with $u^\mu$, we have
\begin{equation}
		u^\mu\nabla^\nu\mathcal{T}_{\mu\nu}=\frac{f_2'(T)\nabla^\nu T}{1+f_2(T)}(\varepsilon-\mathcal{L}_\text{m})=-u^\mu\nabla_\mu\varepsilon-(\varepsilon+p)\nabla_\mu u^\mu.
	\label{mattere}
\end{equation}
If we take the matter Lagrangian density to be
$\mathcal{L}_\text{m}=\varepsilon$\cite{Gron2007,Bertolami2008,Minazzoli2012,Harko2014},
then in cosmological cases,
as discussed in Ref.\cite{Feng2015},
Eq.\eqref{mattere} will return to the usual conservation law of matter in
Friedmann-Lema{\^i}tre-Robertson-Walker(FLRW) metric
$g_{\mu\nu}=\text{diag}(1,-a,-a,-a)$.

In Ref.\cite{Feng2015},
we have constructed two concrete models with
\begin{equation}
	f_1=\begin{cases}\frac{12BH_0^4}{\left|T\right|^{2}}\\\\\frac{BH_0}{\sqrt{\left|T\right|}}\end{cases},\quad f_2=\frac{2AH_0^2}{\Omega_\text{M0}\left|T\right|}\quad  \text{for} \begin{array}{c}\text{Model I}\\\\\text{Model II}\end{array},\nonumber
\end{equation}
where $\Omega_\text{M0}=\Omega_\text{m0}+\Omega_\text{b0}$
is the current density parameter of matter,
and fitted them with SNeIa, CMB, and BAO data.
The best-fit values for the parameters $A,\:B$ are:
\begin{equation}
	\begin{split}
		\text{Model I:}\:A&=0.188\pm0.048,\\
		B&=0.510\pm0.060;\\\\
		\text{Model II:}\:A&=0.633\pm0.012,\\
		B&\text{ is a free parameter}.
	\end{split}
	\label{bestfits}
\end{equation}
We have found that these two models are successful in
describing the observation of the Universe
and its large-scale structure and evolution.

In order to find out whether these two models can pass the Solar system tests,
in the following section,
we are going to consider the gravitation redshift,
geodetic effect, and perihelion precession of these two models.

\section{Spherically symmetric weak field solution}
\label{bgmetric}
Firstly, we calculate the background metric.
Since we are considering the spherically symmetric vacumm solution,
Eq.\eqref{eom} can be rewritten as
\begin{equation}
		\frac4{|e|}f\partial_\beta(|e|S_\sigma^{\:\:\alpha\beta}e_a^{\:\sigma})+4e_a^{\:\sigma}S_\sigma^{\:\:\alpha\beta}\partial_\beta f+4fS_\rho^{\:\:\alpha\sigma}T^\rho_{\:\:\sigma\beta}e_a^{\:\beta}+(1+f_1)Te_a^{\:\alpha}=0,
	\label{eomvac}
\end{equation}
with $f=f(T)=1+f_1(T)+f_1'(T)T$.
Therefore, we are actually dealing with the minimal coupling $f(T)$ theory
since $\mathcal{L}_\text{m}$ vanishes at outside of the central mass.
As mentioned before, one should be careful
when it comes to choosing vierbein fields in spherical coordinates
for $f(T)$ theories.
The general spherically symmetric metric can be written as
\begin{equation}
	g_{\mu\nu}=\text{diag}(\e^{\xi(r,t)},-\e^{\zeta(r,t)},-r^2,-r^2\sin^2\theta).
	\label{metricspherical}
\end{equation}
It is found\cite{Tamanini2012} that the vierbein field
\begin{equation}
	e^a_{\:\mu}=\left(
	\begin{array}{cccc}
		\e^{\frac{\xi}2}&0&0&0\\
		0&\e^{\frac{\zeta}2}\sin\theta\cos\phi&\e^{\frac{\zeta}2}\sin\theta\sin\phi&\e^{\frac{\zeta}2}\cos\theta\\
		0&-r\cos\theta\cos\phi&-r\cos\theta\sin\phi&r\sin\theta\\
		0&r\sin\theta\sin\phi&-r\sin\theta\cos\phi&0
	\end{array}
	\right)
	\label{vierbeinsp}
\end{equation}
is a viable choice of vierbein for $f(T)$ theories
in that it does not impose extra constraints on the form of $f(T)$
and it preserves the Birkhoff's theorem, that is
\begin{equation}
	\xi=\xi(r),\quad\zeta=\zeta(r).
	\label{xizeta}
\end{equation}
The torsion scalar is then given by
\begin{equation}
	T(r)=\frac2{r^2}\e^{-\zeta}(1+\e^{\frac{\zeta}2})(1+\e^{\frac{\zeta}2}+r\xi'),
	\label{tscalar}
\end{equation}
where the prime indicates derivative with respect to $r$.

Thus, Eq.\eqref{eomvac} can reduce to
\begin{equation}
	\begin{split}
		f(T)[\xi'(r)+\zeta'(r)]=&2[1+\e^{\frac{\zeta(r)}2}]f'(T)T'(r),\\
		\e^{\zeta(r)}r^2T^2f_1'(T)=&2f(T)[\e^{\zeta(r)}-1-r\xi'(r)].
	\end{split}
	\label{vacumm1}
\end{equation}
In the weak field limit $\xi,\zeta\ll1$,
these equations can be solved up to linear approximation as
\begin{equation}
	\begin{split}
		\xi_\text{I}(r)=&-\frac{2M}r+\frac{ 3BH_0^4r^4}{20},\\
		\zeta_\text{I}(r)=&\frac{2M}r-\frac{9BH_0^4r^4}{10}
	\end{split}
	\label{vacummsolI}
\end{equation}
for Model I, and
\begin{equation}
	\begin{split}
		\xi_\text{II}(r)=&-\frac{2M}r+\frac{BH_0r}{\sqrt{2}},\\
		\zeta_\text{II}(r)=&\frac{2M}r
	\end{split}
	\label{vacummsolII}
\end{equation}
for Model II,
where the integral constant is chosen to be
the central spherically symmetic mass $M$
as in the Schwarzschild solution.
These results can also be derived from
the general weak field solution for minimal coupling power law
$f(T)$ theory given in Ref.\cite{Ruggiero2015}.

With the (vacumm) background vierbein field solved,
we are ready to calculate the Solar system effects.

\section{Gravitation redshift}
\label{gr}
The gravitation redshift for a radiation signal is given by
\begin{equation}
	\frac{\nu}{\nu_0}=\frac{e_0^{\:0}(r)}{e_0^{\:0}(r_0)},
	\label{redshift}
\end{equation}
where $\nu,\nu_0$ are the frequencies of the signal measured at the position of $r,r_0$, respectively.
Since this effect does not involve any massive test particle,
%(the coupling between radiation and gravity is assumed to be minimal for simplicity)
the background solutions \eqref{vacummsolI} and \eqref{vacummsolII} can be used directly.
Thus, up to linear approximation, the gravitation redshift is
\begin{equation}
	\left(\frac{\nu}{\nu_0}\right)_\text{I}\simeq1-\frac Mr+\frac M{r_0}+ \frac{3BH_0^4}{40}(r^{4}-r_0^{4})
	\label{redshiftI}
\end{equation}
for Model I and
\begin{equation}
	\left(\frac{\nu}{\nu_0}\right)_\text{II}\simeq1-\frac Mr+\frac M{r_0}+ \frac{BH_0}{2\sqrt{2}}(r-r_0)
	\label{redshiftII}
\end{equation}
for Model II.
The common terms in each case are the prediction of GR,
and the last terms are the modifications from the models.

The Hydrogen-Maser in Gravity Probe A redshift experiment\cite{Vessot1980}
launched a spacecraft nearly vertically upward to 10000 km
to measure this effect.
Using the cosmologically best-fit values \eqref{bestfits},
we have the gravitation redshift modifications
$\delta(\frac{\nu}{\nu_0})_\text{I}\le9.045\times10^{-78}$ for Model I and
$\delta(\frac{\nu}{\nu_0})_\text{II}=2.331B\times10^{-20}$ for Model II.
The redshift experiment reaches a
$\delta_\text{rs}=10^{-14}$ accuracy\cite{Vessot1980}.
It is obvious that the modifications are within the error for both models
if $B<10^6$ for Model II.

\section{Geodetic effect}
\label{ge}
Consider a gyroscope described by a spin 4-vector $s^\mu$
orbiting around the central mass.
Its 4-velocity is given by $u^\mu=(u^t,0,0,u^\phi)$
where $u^\phi$ is the orbiting angular velocity,
and $u_\mu u^\mu=1$.
And $s^\mu$ is spacelike such that $u_\mu s^\mu=0$.
It is found that as long as the direct effects of tidal forces can be neglected,
$s^\mu$ is Fermi-Walker transported along $u^\mu$\cite{MTW1973,Will2014},
that is
\begin{equation}
	u^\nu\nabla_\nu s^\mu=u^\mu(u^\lambda\nabla_\lambda u^\nu)s_\nu.
	\label{FWtrans}
\end{equation}
Contracting Eq.\eqref{covder} with the induced metric $h^{\mu\lambda}=g^{\mu\lambda}-u^\mu u^\nu$,
we have
\begin{equation}
		h^{\mu\lambda}\nabla^\nu\mathcal{T}_{\mu\nu}=h^{\mu\lambda}\nabla_\mu p-(\varepsilon+p)u^\mu\nabla_\mu u^\lambda=-h^{\mu\lambda}\frac{f_2'(T)\nabla_\mu T}{1+f_2(T)}(p+\mathcal{L}_\text{m}).
	\label{force1}
\end{equation}
Combining Eqs.\eqref{FWtrans} and \eqref{force1},
and supressing the hydrodynamical term $\nabla_\nu p$
which is not related to $f_2$, we have
\begin{equation}
	\begin{split}
		u^\nu\nabla_\nu s^\mu=&\frac{\diff s^\mu}{\diff t}+\Gamma^\mu_{\nu\lambda}s^\lambda u^\nu=s^\lambda u^\mu\frac{f_2'(T)\partial_\lambda T}{1+f_2(T)},
	\end{split}
	\label{force2}
\end{equation}
where $\Gamma^\mu_{\nu\lambda}$ is the Levi-Civit{\'a} connection
and we have taken $\mathcal{L}_\text{m}=\varepsilon$\cite{Gron2007,Bertolami2008,Minazzoli2012,Harko2014}.
The $r,\theta,\phi$ components of Eq.\eqref{force2} are
\begin{equation}
	\begin{split}
		\frac{\diff s^r}{\diff t}&+\frac12\xi'(r)\e^{\xi-\zeta}s^tu^t-r\sin^2\theta\e^{-\zeta}s^\phi u^\phi=0,\\
		\frac{\diff s^\phi}{\diff t}&+\frac1rs^ru^\phi+\cot\theta s^\theta u^\phi=u^\phi s^r\frac{f_2'(T)T'(r)}{1+f_2(T)},\\
		\frac{\diff s^\theta}{\diff t}&-\sin\theta\cos\theta s^\phi u^\phi=0.
	\end{split}
	\label{eqsgp1}
\end{equation}
The first and third equations vanish because $u^r,u^\theta=0$.
Since $s^\mu u_\mu=0$, let $\theta=\pi/2$, we can put
\begin{equation}
	s^tu^t=\e^{-\xi}r^2s^\phi u^\phi
	\label{st}
\end{equation}
into Eq.\eqref{eqsgp1} and then have
\begin{equation}
	\begin{split}
		\frac{\diff s^r}{\diff t}&+\left[ \frac12\xi'(r)r^2-r \right]\e^{-\zeta}s^\phi u^\phi=0,\\
		\frac{\diff s^\phi}{\diff t}&+\left[ \frac1r-\frac{f_2'(T)T'(r)}{1+f_2(T)} \right]s^ru^\phi=0,\\
	\frac{\diff s^\theta}{\diff t}&=0.
	\end{split}
	\label{eqsgp2}
\end{equation}
The general solution of Eq.\eqref{eqsgp2} can be written as
\begin{equation}
	\begin{split}
		s^r&=s^r_0\sin\Omega t,\\
		s^\phi&=s^\phi_0\cos\Omega t,\\
		s^\theta&=s^\theta_0=\text{const.},
	\end{split}
	\label{solgp}
\end{equation}
where the angular frequency $\Omega$ is given by
\begin{equation}
	\Omega^2=(u^\phi)^2\left[r- \frac12\xi'(r)r^2 \right]\left[ \frac1r-\frac{f_2'(T)T'(r)}{1+f_2(T)} \right]\e^{-\zeta}.
	\label{omega}
\end{equation}

On the other hand,
from Eq.\eqref{force1}, we have
\begin{equation}
	u^\nu\nabla_\nu u^\mu=\frac{\diff u^\mu}{\diff t}+\Gamma^\mu_{\nu\lambda}u^\nu u^\lambda=h^{\mu\nu}\frac{f_2'(T)\nabla_\nu T(r)}{1+f_2(T)}.
	\label{geo}
\end{equation}
Using $u_\mu u^\mu=1$, we have
\begin{equation}
	(u^\phi)^2=\frac{\frac12\xi'(r)-\frac{f_2'(T)T'(r)}{1+f_2(T)}}{r(1-\frac r2\xi'(r))}.
	\label{uphi}
\end{equation}
Thus, the precessing rate of the gyroscope $\phi_\text{ge}=u^\phi-\Omega$.
For the two models being considered, up to linear approximation, it is
\begin{equation}
	\begin{split}
		\phi_\text{ge,I}=&\sqrt{\frac M{r^3}}\left( \frac{3M}{2r}-\frac{AH_0^2r^2}{8\Omega_\text{M0}}-\frac{3BH_0^4r^4}{40} \right),\\
		\phi_\text{ge,II}=&\sqrt{\frac M{r^3}}\left( \frac{3M}{2r}-\frac{AH_0^2r^2}{8\Omega_\text{M0}}+\frac{5BH_0r}{8\sqrt{2}} \right).
	\end{split}
	\label{phige}
\end{equation}
The first term $3M/2r$ is the prediction of GR,
the rest terms are the modifications from the models.
Thus the relative modifications are
\begin{equation}
	\begin{split}
		\Delta\phi_\text{ge,I}=&-\frac{AH_0^2r^3}{12\Omega_\text{M0}M}-\frac{BH_0^4r^5}{20M},\\
		\Delta\phi_\text{ge,II}=&-\frac{AH_0^2r^3}{12\Omega_\text{M0}M}+\frac{5BH_0r^2}{12\sqrt{2}M}.
	\end{split}
	\label{phigerel}
\end{equation}

The Gravity Probe B mission\cite{Everitt2011} tested this effect
with 4 gyroscopes at the typical altitude of 642 km.
Analysis of the data resulted in a relative error of geodetic precessing rate
less than $0.4\%$.
With the cosmologically best-fit value \eqref{bestfits} and
the mass of Earth ($GM_\oplus\simeq22487.9\:\text{eV}^{-1}$, with the gravitational constant $G$ recovered),
the relative modifications of our models are
$\left|\Delta\phi_\text{ge,I}\right|\le3.005\times10^{-31}$ and
$\Delta\phi_\text{ge,II}=2.152B\times10^{-11}$.
It is obvious that the modifications from both models are within the experiment error
if $B<10^8$ for Model II.

\section{Perihelion precession}
\label{pp}
For a massive test particle orbiting around the central mass,
the equation of motion can be retained from Eq.\eqref{geo}:
\begin{equation}
	\frac{\diff^2x^\mu}{\diff\tau^2}+{\Gamma}^\mu_{\alpha\beta}\frac{\diff x^\alpha}{\diff\tau}\frac{\diff x^\beta}{\diff\tau}=h^{\mu\nu}\frac{f_2'(T)\nabla_\nu T(r)}{1+f_2(T)}.
	\label{geo1}
\end{equation}
Assume that the orbit lies in the $\theta=\pi/2$ plane and $\diff\theta=0$.
The $t$ and $\phi$ components of the equation of motion are
\begin{equation}
	\begin{split}
		\frac{\diff^2t}{\diff\tau^2}&=-\left[ \xi'(r)+\frac{f_2'(T)T'(r)}{1+f_2(T)} \right]\frac{\diff r}{\diff\tau}\frac{\diff t}{\diff\tau},\\
		\frac{\diff^2\phi}{\diff\tau^2}&=-\left[ \frac2r+\frac{f_2'(T)T'(r)}{1+f_2(T)} \right]\frac{\diff r}{\diff\tau}\frac{\diff\phi}{\diff\tau},
	\end{split}
	\label{geo2}
\end{equation}
which can be integrated as
\begin{equation}
	\frac{\diff t}{\diff\tau}=\frac{k\e^{-\xi}}{1+f_2},\quad\frac{\diff\phi}{\diff\tau}=\frac{l}{r^2(1+f_2)},
	\label{geo3}
\end{equation}
where $k,h$ are integral constants.

Now from the line element
\begin{equation}
		(\diff\tau)^2=\e^\xi(\diff t)^2-\e^\zeta(\diff r)^2-r^2(\diff\phi)^2,
	\label{ppmetric}
\end{equation}
we have
\begin{equation}
	\begin{split}
		\left(\frac{\diff r}{\diff\tau}\right)^2=&\e^{\xi-\zeta}\left(\frac{\diff t}{\diff\tau}\right)^2-r^2\e^{-\zeta}\left(\frac{\diff\phi}{\diff\tau}\right)^2\\
		=&\frac{k^2\e^{-\xi-\zeta}}{(1+f_2)^2}-\frac{l^2\e^{-\zeta}}{r^2(1+f_2)^2}.
	\end{split}
	\label{ppdrdt}
\end{equation}
Up to linear approximation,
Eq.\eqref{ppdrdt} can be expressed as
\begin{equation}
	\begin{split}
		\left( \frac{\diff r}{\diff\tau} \right)_\text{I}^2=&k^2-1+\frac{2M}r+\frac{l^2}{r^2}\left( \frac{2M}r-1 \right)\\
		&+\frac{3 B H_0^4}{4}\left( k^2-\frac65 \right)r^4-\frac{9 B H_0^4 l^2 r^2}{10}\\
		&-\frac{A H_0^2 k^2 r^2}{2\Omega_\text{M0}}+\frac{A H_0^2 l^2}{2\Omega_\text{M0}}
	\end{split}
	\label{ppeqI}
\end{equation}
for Model I and
\begin{equation}
	\begin{split}
		\left( \frac{\diff r}{\diff\tau} \right)_\text{II}^2=&k^2-1+\frac{2M}r+\frac{l^2}{r^2}\left( \frac{2M}r-1 \right)\\
		&-\frac{ B H_0 k^2 r}{\sqrt2}-\frac{A H_0^2 k^2 r^2}{2\Omega_\text{M0}}+\frac{A H_0^2 l^2}{2\Omega_\text{M0}}
	\end{split}
	\label{ppeqII}
\end{equation}
for Model II.
The first line of each model represents the GR case.
It can be compared to the classical equation of motion
\begin{equation}
	\frac12m_0\dot{r}^2+V+\frac{L}{2m_0r^2}=E,
	\label{classicaleq}
\end{equation}
where $V,E$ are the potential and total energy,
and $L$ is the angular momentum,
and $m_0$ is the mass of the test particle.
In our cases, both $L$ and $E$ are constants.
Thus in the first lines of Eqs.\eqref{ppeqI} and \eqref{ppeqII}, we can identify
$V_\text{Newt.}=-\frac{M}r$ as the Newtonian potential,
$\mathcal{E}=\frac{k^2-1}2$ as the conserved energy per unit mass,
$l$ as the conserved angular momentum per unit mass,
and $V_\text{GR}=-\frac{l^2M}{r^3}$ as the potential from GR.
The rest lines of Eqs.\eqref{ppeqI} and \eqref{ppeqII} can be seen as
additional potentials from the modifications of the models.
Thus, the equation of motion can be re-expressed as
\begin{equation}
	\frac12\left( \frac{\diff r}{\diff\tau} \right)^2=\mathcal{E}+\frac{l^2}{2r^2}-V_\text{Newt.}-V_\text{GR}-V_\text{mod.},
	\label{classicalid}
\end{equation}
with $V_\text{mod.}=V_1+V_2+V_3+V_4$ and
\begin{equation}
	\begin{split}
		V_1(r)=&\left( 2\mathcal{E}+1 \right)\frac{A H_0^2 r^2}{4\Omega_\text{M0}},\\
		V_2=&-\frac{A H_0^2 l^2}{4\Omega_\text{M0}}
	\end{split}
	\label{potentials0}
\end{equation}
for both models and
\begin{equation}
	\begin{split}
		V_{3,\text{I}}(r)=&-\frac{3 B H_0^4}{8}\left( 2\mathcal{E}-\frac15 \right)r^4,\\
		V_{4,\text{I}}(r)=&\frac{9 B H_0^4 l^2 r^2}{20}
	\end{split}
	\label{potentialsI}
\end{equation}
for Model I and
\begin{equation}
	\begin{split}
		V_{3,\text{II}}(r)=&\frac{ B H_0 \left( 2\mathcal{E}+1 \right) r}{2\sqrt2},\\
		V_{4,\text{II}}(r)=&0
	\end{split}
	\label{potentialsII}
\end{equation}
for Model II.

$V_2$ and $V_{4,\text{II}}$ are constants,
which do not contribute to the precession.
All the other potentials are power law functions of $r$.
For a potential $V(r)$ in the form $V(r)=\gamma r^\lambda$,
it is found\cite{Xu2011} that the extra perihelion precession per orbital period is
\begin{equation}
	\delta\phi=\frac{\lambda\gamma\left( 1-e_\text{ecc.}^2 \right)^{\lambda+1}r_a^{\lambda+1}}{e_\text{ecc.}M}I\left( e_\text{ecc.},\lambda \right),
	\label{addpp}
\end{equation}
where $e_\text{ecc.}$ and $r_a$ are the eccentricity and the semi major axis of the orbit, respectively, and
\begin{equation}
	I\left( e_\text{ecc.},\lambda \right)=\int_0^{2\pi}\frac{\cos\theta\diff\theta}{\left( 1+e_\text{ecc.}\cos\theta \right)^{\lambda+1}}.
	\label{ielambda}
\end{equation}
For a small eccentricity $e_\text{ecc.}$, $I(e_\text{ecc.},\lambda)\simeq-e_\text{ecc.}\pi(1+\lambda)$.
Hence, the modification of the perihelion precession from GR is
\begin{equation}
	\delta\phi_\text{GR}\simeq\frac{6l^2\pi}{\left( 1-e_\text{ecc.}^2 \right)^2r_a^2},
	\label{ppGR}
\end{equation}
which is the well known result.
The extra modifications of perihelion precession are
\begin{equation}
		\delta\phi_\text{I}=\left[ \frac{A H_0^2}{\Omega_\text{M0}}\left( 1+2\mathcal{E} \right)+\frac{9 B H_0^4 l^2}{5} \right]\frac{3\left( e_\text{ecc.}^2-1 \right)^3\pi r_\text{a}^3}{2M}-\frac{3B H_0^4\left( e_\text{ecc.}^2-1 \right)^5\left( 10\mathcal{E}-1 \right)\pi r_\text{a}^5}{2M}
	\label{ppmodI}
\end{equation}
for Model I and
\begin{equation}
		\delta\phi_\text{II}=\frac{3AH_0^2\left( e_\text{ecc.}^2-1 \right)^3\left( 1+2\mathcal{E} \right)\pi r_\text{a}^3}{2M\Omega_\text{M0}}-\frac{BH_0\left( e_\text{ecc.}^2-1 \right)^2\left( 2\mathcal{E}+1 \right)\pi r_\text{a}^2}{\sqrt{2}M}
	\label{ppmodII}
\end{equation}
for Model II.

The orbit of Mercury has a semi major axis of $r_\text{a}=5.791\times10^{7}\:\text{km}$,
eccentricity of $e_\text{ecc.}=0.206$,
and an orbital period of $T_0=87.969\:\text{d}$.
With the cosmologically best-fit values of $A,\:B$ \eqref{bestfits}
and the Solar mass ($GM_\odot\simeq7.483\times10^9\:\text{eV}^{-1}$, with the gravitational constant recovered),
we can calculate that the extra precession rates are
$\left|\delta\phi_\text{I}/T_0\right|<\mathrm{O}(10^{-14})\:\text{mas/yr}$ for Model I and
$\delta\phi_\text{II}/T_0=-26.102B\:\text{mas/yr}$ for Model II.
The perihelion precession of Mercury measured from experiments
matches the GR's prediction within an error of $\delta\phi_\text{E}/T_0=0.43\:\text{mas/yr}$\cite{Nordtvedt2000}.
Thus, the modification from Model I is within the experiment error.
However, the perihelion precession requires $B<1.647\times10^{-2}$ for Model II.

\section{Conclusion and discussion}
\label{sum}
In this paper, we have studied the gravitation redshift,
geodetic effect and perihelion precession in the two concrete models
of $f(T)$ theory with nonminimal torsion-matter coupling extension,
which have been previously constructed and investigated for cosmology.
It is found that Model I can pass all three Solar system tests.
For Model II, the Solar system tests request the parameter $B<1.647\times10^{-2}$,
otherwise the theoretical value will surpass the bound of measure for the perihelion precession of Mercury.

Neither the gravitation redshift nor Shapiro time delay involves massive particle.
Since the coupling between radiation and torsion is still minimal in Models I and II,
these effects are practically the same as the minimal coupling theory\cite{Iorio2012,Xie2013,Farrugia2016}.
Furthermore, the test of Shapiro time delay has been omitted in this paper
because its constraint is looser than that of the gravitation redshift
for the parameters $A$ and $B$ in the Models I and II.

In a nutshell,
we have shown that Models I and II successfully describe
not only the observation of the Universe and its large-scale structure and evolution,
but also the Solar system effects of gravitation.
Therefore, these models are realistic ones.
However, we find no acceptable $H_0$ and $\Omega_\text{m0}$
when $f_1(T)$ and $f_2(T)$ are both propotional to $T^2$.
Further research is needed as to whether or not there is a better $f(T)$ model
with nonminimal torsion-matter coupling extension.
For instance, one can consider the case of $f_1(T)\propto T^{m}$
and $f_2(T)\propto T^{n}$ and take optimum of $(m,n)$
using observation at the large and small scales.
We will consider this question in the future work.

\acknowledgments
This work is partially supported by National Science Foundation of China grant No.~10671128 and Key Project of Chinese Ministry of Education grant, No.~211059.

\bibliography{ref}
\bibliographystyle{apsrev}
\end{document}